\documentstyle[psfig]{article}
\begin{document}
\begin{center} 
{\bf Auto-correlation of Binary Stars}
\end{center} 
\vspace{1cm}

\begin{center}
S. K. Saha\footnote{e-mail: sks@iiap.ernet.in}
and 
D. Maitra 
\footnote{Summer project
student from Indian Institute of Technology Kanpur 208016, India} \\
Indian Institute of Astrophysics\\
Bangalore 560034\\ 
India
\end{center}
\vspace{1cm}

\noindent
{\bf Abstract}: Speckle interferometric technique is used to record a
series of short exposure images of several close binary stars with sub-arcsecond
separation through a narrow band filter centred at H$\alpha$ at the
Cassegrain focus of the 2.34 meter (m) Vainu Bappu telescope (VBT), situated at
Vainu Bappu Observatory (VBO), Kavalur . The  auto-correlation method is
developed under Image Reduction Analysis Facility (IRAF).
Wiener filter is included in the programme to eliminate spurious high frequency 
contributions; a few sets of data provide the optimised results. The 
auto-correlated image of these stars gives the separation of the binary 
components. 
\vspace{0.3cm}

\vspace{0.3cm}

\vspace{0.3cm}

\begin{center} 
{\bf 1. Introduction} 
\end{center} 
\vspace{0.3cm}

\noindent
The speckle interferometric technique [1] is widely used
to decode the deleterious effect of the atmospheric turbulence that limits
the sensitivity of ground-based telescopes. Recent reviews by Saha
[2, 3] describe in depth the utility of this method. One of the 
important subjects in stellar astrophysics is the
studies of close binary stars, that play a fundamental role in measuring
stellar masses, providing a benchmark for stellar evolution calculations; a
long-term benefit of interferometric imaging is a better calibration of the
main-sequence mass-luminosity relationship [4]. The procedure
in obtaining masses of stars involves combining the spectroscopic orbit with the
astrometric orbit, which is the projection of true orbit on the sky, from the
interferometric data [5]. In order to derive accurate orbital
elements and masses, luminosities and distances, continuous observations
of spectroscopic binaries are essential. Major contributions in resolving close
binary systems, among others, came from the Center for High Angular Resolution 
Astronomy (CHARA) at Georgia State University, USA [6-9]. Survey of visual 
and interferometric binary stars have also been reported 
by the other groups [10-19].
\vspace{0.3cm}

\noindent
Most of these results that appeared in various journals till date are from the
usage of the auto-correlation technique though schemes, like Knox-Thomson
algorithm [20], shift and add [21], triple correlation [22], and Blind Iterative
deconvolution (BID) technique [23] restore 
the degraded images. All these schemes, barring BID, depend on the statistical 
treatment of a large number of images. Often, it may not be possible
to record a large number of images within the time interval over which the
statistics of the atmospheric turbulence remains stationary. In such cases,
where only a few images are available, we describe an auto-correlation algorithm
that is developed under IRAF; this algorithm has been applied on degraded
images of several binaries, obtained at the 2.34~m VBT at Kavalur. 
\vspace{0.3cm}

\begin{center}
{\bf 2. Outline of the theory of speckle interferometry}
\end{center}
\vspace{0.3cm}

\noindent
The instantaneous image intensity distribution $I({\bf x})$ 
recorded by the detector (CCD/photographic plate/eye etc.) is actually a
convolution of the instantaneous `Point Spread Function (PSF)'
$S({\bf x})$, with the actual intensity distribution of the object  
$O({\bf x})$. Here ${\bf x}$ is the two dimensional position vector. The PSF 
describes how light from a point source (a two dimensional $\delta$ function on 
the sky) gets distributed across the detector. The variability of the
corrugated wave-front yields 'speckle boiling' and is the source
of speckle noise that arises from difference in registration
between evolving speckle pattern and the boundary of the PSF area
in the focal plane. These specklegrams have additive noise
contamination, $N({\bf x})$, which includes all additive
measurement of uncertainties. This may be in the form of (i)
photon statistics noise, and (ii) all distortions from the
idealized isoplanatic model represented by the convolution of
$O({\bf x})$ with $S({\bf x})$ that includes non-linear geometrical distortions.
\vspace{0.3cm}

\noindent
If instead of dealing in the real space
we choose to work in Fourier space, our life becomes simpler. This is because 
the operation of convolution in real space becomes simple multiplication in the 
Fourier space. Mathematically speaking:

\begin{equation}
I({\bf x}) = O({\bf x})\ast S({\bf x}) + N({\bf x}),
\end{equation}

\noindent
where `$\ast$' signifies convolution operation.

\noindent
In {\it Fourier} domain, equation (1) can be written as:

\begin{equation}
I({\bf u}) = O({\bf u})\cdot S({\bf u}) + N({\bf u}),
\end{equation}

\noindent
where ${\bf u}$ is the two dimensional spatial frequency vector and `$\cdot$'
implies the ordinary operation of multiplication. The quantity $O({\bf u})$ is
called the {\it Object Spectrum}, and $S({\bf u})$ the {\it Transfer function}.
\vspace{0.3cm}

\noindent
If the exposure time is short enough then the atmosphere can be taken to be 
static during the exposure. The common term that describes this is that `the 
atmosphere has
been {\it freezed} during the exposure'. From experiments it has been found  
that for exposures below 20~ms the atmosphere can be considered as fairly static
for reasonably stable nights.
\vspace{0.3cm}

\noindent
A large number of such short exposure images (specklegrams) of the object of 
interest is taken. Also images of a reference star which is known to be a 
unresolved star 
(i.e., a point source, a two dimensional $\delta$ function on the sky)
and which lies in the same isoplanatic patch as that of the program star, is 
taken at the same time. It is assumed that the atmospheric distortion affects
the reference star in the same way as it does to the program star.
\vspace{0.3cm}

\noindent
The ensemble average of the power spectrum is given by

\begin{equation}
<{\mid I({\bf u})\mid}^{2}> = {\mid O({\bf u})\mid}^{2}
\cdot <{\mid S({\bf u})\mid}^{2}> + <{\mid N({\bf u})\mid}^{2}>.
\end{equation}

\noindent
The reference star being a single star, the Fourier transform of its image 
is unity, i.e., $O_{ref}({\bf u})$ = 1.
\vspace{0.3cm}

\noindent
The power spectrum of the program star is therefore given by

\begin{equation}
\mid O_{pro}({\bf u})\mid^{2}=\frac{<{\mid I_{pro}({\bf u})\mid}^{2}>}
{<{\mid I_{ref}({\bf u})\mid}^{2}>},
\end{equation}

\noindent
and by Wiener-Kinchin theorem, the autocorrelation of the object is given by

\begin{equation}
A[O({\bf x})]={\mathcal FT}^{-1}[{\mid O({\bf u})\mid}^{2}],
\end{equation}

\noindent
where $A$ denotes Autocorrelation and ${\mathcal FT}^{-1}$
denotes the operation of inverse Fourier transform.

\begin{center} 
{\bf 3. Data Processing and Noise Reduction}
\end{center}
\vspace{0.3cm}

\noindent
The program for finding the power spectrum is written in FORTRAN. It uses
IRAF commands and hence the macros in IRAF are called and used. The program 
is written in such a way that a large number of data files can be handled and
large number of images can be averaged to increase the signal-to-noise ratio.
The set of input specklegrams of object and the reference stars are read and
their power spectrum is calculated. The object power spectrum and the image 
power spectrum are averaged in the Fourier space. Finally the
autocorrelated image is calculated using another programme.
\vspace{0.3cm}

\noindent 
The disadvantage with a division as in equation (4), is that the zeros in the 
denominator will corrupt the ratio and spurious high frequency components
will be created in the reconstructed image. Moreover, a certain amount of noise 
is inherent in any kind of observation. In our present case noise is
primarily due to thermal electrons in the CCD interfering with the signal. Most of this noise is in the high spatial frequency regime.
\vspace{0.3cm}

\noindent
In order to get rid of this high frequency noise as much as
possible, a Wiener filter was used. Applying the Wiener filter is
essentially the process of convolving the noise degraded image
$I({\bf x})$ with the Wiener filter. The Wiener filter is applied in the 
frequency domain. Then the original image spectrum ${I}({\bf u})$ is estimated 
from the degraded image spectrum $I^\prime({\bf u})$ by simply multiplying the
image spectrum with the Wiener filter $W({\bf u})$. However, this will reduce 
the resolution in the reconstructed image. The advantage is the spurious higher 
frequency contribution are eliminated.

\begin{equation}
{I}({\bf u}) = I^\prime({\bf u})\cdot W({\bf u}).
\end{equation}

\noindent
The Wiener filter, in the frequency domain, has the following functional form:

\begin{equation}
W({\bf u}) = \frac{H^{*}({\bf u})}{{\mid H({\bf u}) \mid}^{2}+\frac{P_{n}({\bf 
u})}{P_{s}({\bf u})}}.
\end{equation}

\noindent
where, $H({\bf u})$ is the Fourier transform of the point-spread-function,
$P_{s}({\bf u})$ the power spectrum of signal process, and
$P_{n}({\bf u})$ the power spectrum of noise process.

\noindent
The term $P_{n}({\bf u})$/$P_{s}({\bf u})$ can be interpreted as
reciprocal of signal to noise ratio. In our case, the noise is
due to the CCD. We developed an IRAF-based algorithm, where a
Wiener parameter, $w$, is added to the PSF power spectrum in order
to avoid zeros in the PSF power spectrum that helps in
reconstructions with a few frames. The classic Weiner filter that
came out of the electronic information theory where
diffraction-limits do not mean much, is meant to deal with signal
dependent 'coloured' noise. In practice, this term is usually just
a constant, a 'noise control parameter' whose scale is estimated
from the noise power spectrum. In this case, it assumes that the
noise is white and that one can estimate its scale in regions of
the power spectrum where the signal is zero (outside the diffraction-limit for 
an imaging system). The expression for the Wiener filter simplifies to:

\begin{equation}
I({\bf u})=\frac{P_{ref}({\bf u})}{P_{ref}({\bf u})+w},
\end{equation}

\noindent
where, $w$ is the noise-variance and termed as Wiener filter
parameter in the program. 
\vspace{0.3cm}

\noindent
To get an optimally autocorrelated image, a judicious choice of the Wiener 
filter parameter is made according to the procedure described below:
\vspace{0.3cm}

\noindent
For a very wide range of Wiener filter parameter values, the
autocorrelated images are constructed. A small portion
(16 X 16 pixels) of each image, far from the centre, is sampled to find-out the 
standard deviation in the intensity values of the pixels, $\sigma_{noise}$.
The plot of standard deviation thus obtained against the Wiener filter
parameter (Figures 1(c)) shows a minimum. The abscissa corresponding to
this minimum gives the optimum Wiener filter parameter value.
\vspace{0.3cm}

\noindent
The nature of $\sigma_{noise}$ vs. WFP plot is understood as follows:\\
The noise in our data is primarily in the high frequency 
region whereas the signal is at a comparatively low frequency. With zero WFP, 
there is no filtering and hence there is ample noise. As the WFP value is 
gradually increased from zero, more and more high frequency noise is cut off 
and the $\sigma_{noise}$ goes down. It attains a minimum when WFP value is 
just enough to retain the signal and discard the higher frequency noise.
However, at higher WFP region, we are over compensating for the
signal, and therefore blurring of the image starts to occur. This leads 
to the sharp increase in $\sigma_{noise}$. The `ringing' effect due to sharp 
cutoff comes into play.
\vspace{0.3cm}

\noindent
Computer simulations were carried out by convolving ideal star
images having Gaussian profile with a random PSF to generate
speckle pattern. The plot of $\sigma_{noise}$ vs. WFP also gives
similar results though some of the major sources of noise, e.g.,
thermal noise due to electron motion in the CCD, effect due to
cosmic rays on the frame, etc. were not incorporated in the simulation.
\vspace{0.3cm}

\begin{center} 
{\bf 4. Instrumentation} 
\end{center}
\vspace{0.3cm}

\noindent
The 2.34~m VBT has two accessible foci for back end instrumentation $-$ a 
prime focus (f/3.25 beam) and a Cassegrain focus (f/13 beam). The latter was 
used for the observations described in this paper. The Cassegrain focus has 
an image scale of 6.7 arcseconds per mm. This was further magnified using a 
microscope objective [24]. This enlarged image was recorded 
through a 5~nm filter centred at H$\alpha$ using an EEV intensified CCD camera 
which provides a standard CCIR video output of the recorded scene. The 
interface between the intensifier screen and the CCD chip is a fibre-optic 
bundle which 
reduces the image size by a factor of 1.69. A DT-2851 frame grabber card 
digitises the video signal. This digitiser re samples the pixels of each row 
(385 CCD columns to 512 digitized samples) and introduces a net reduction in the
row direction by a factor of 1.27. Finally an image-scale of
0.015 arcsecond per pixel is achieved.
The frame grabber can store upto two images on its onboard memory. These
images are then written onto the hard disc of a computer.
\vspace{0.3cm}

\begin{center} 
{\bf 5. Results} 
\end{center}
\vspace{0.3cm}

\noindent
The separations of HR4689 and HR5858, observed on 28 February 1997 were 
estimated using the procedure described above. The average seeing
was around 1.5 arcsec. Figures 1a, 1b, 1c show the 
specklegrams of HR4689, minimum noise autocorrelated image of HR4689,
and variation of noise with Wiener filter parameter respectively. The 
separation of HR4689 was found to be 0.17 $\pm$ 0.03 arcseconds and 
that of HR5858 was 0.23 $\pm$ 0.03 arcseconds. The separations observed are 
compatible with the values published in the CHARA catalogue [25].

\begin{figure}[h]
\centerline{
\psfig{figure=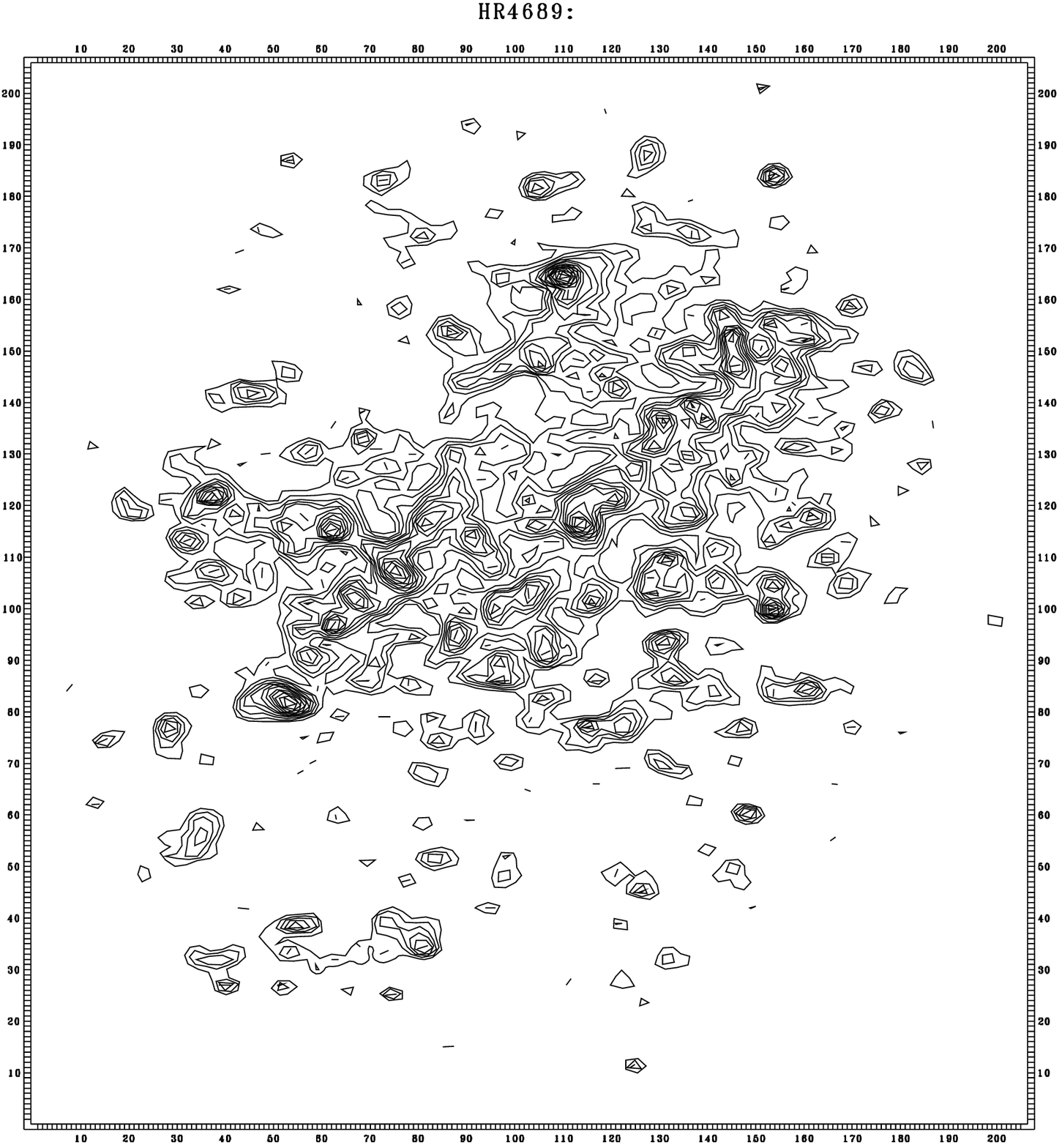,height=5.0cm,angle=270}
\psfig{figure=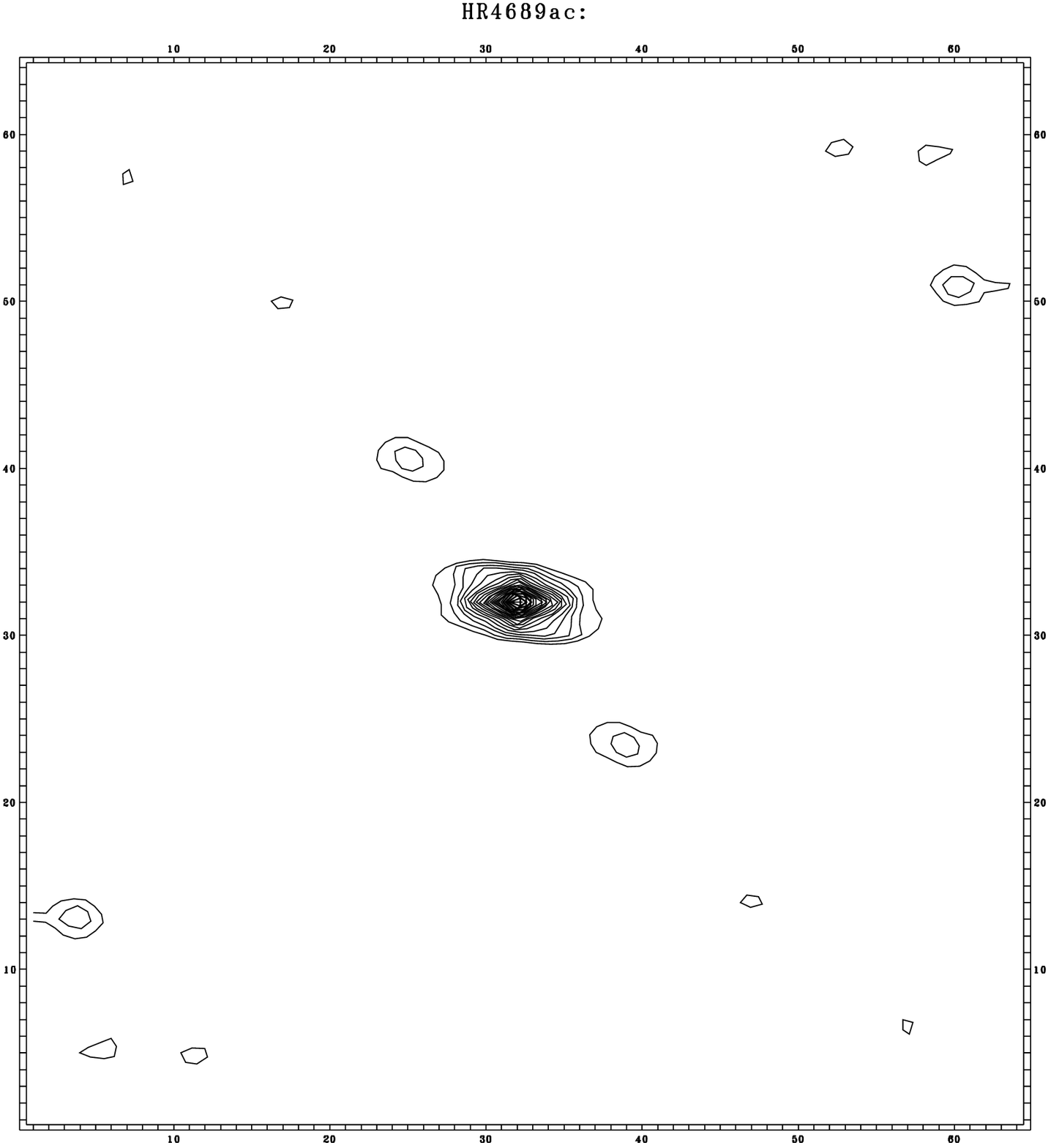,height=5.0cm,angle=270}
}
\hspace{3.75cm} (a) \hspace{3.5cm} (b)
\vspace{0.5cm}

\centerline{
\psfig{figure=wf_stddev.epsi,height=7.5cm,angle=-90}
}
\centerline{(c)}
\caption{: 1a, 1b, 1c show the specklegram of HR4689, the autocorrelation of
HR4689 and $\sigma_{noise}$ vs. WFP plot respectively. The axes of the figures
1(a) and 1(b) are the pixel value; each pixel value is 0.015 arc-seconds. Since 
the inherent property of the autocorrelation method is to produce double images 
with 180$^\circ$ ambiguities of a binary source, one of the two contours on 
either side of the central one (fig 1b) is the secondary component; the central
contours represents the primary component. The contours at the corners are the 
artifacts.
}
\end{figure}

\begin{center} 
{\bf 6. Discussion and conclusions} 
\end{center} 
\vspace{0.3cm}

\noindent
The program of speckle imaging of binary stars at VBT, Kavalur has been a
successful one. Several experiments and computer simulations have been carried 
out for improving the speckle interferometer at VBT [3]. 
Algorithms are in stage of development for Fourier phase recovery too. Noise 
in the data is reduced to an optimum level with the judicious use of Wiener 
filter. Recent developments in CCD technology and associated photon counting 
devices of high quantum efficiency promise better signal to noise ratio, and 
hence better accuracy in determining the orbital parameters of
binaries. However, in this algorithm, it gives a better signal to
noise ratio if the reference star is brighter than the program
star, both being in the same isoplanatic patch. This is very
difficult to achieve in reality and when it is not fulfilled, we
have tried to make a compensation by increasing the number of
frames for the reference.
\vspace{0.3cm}

\noindent
{\bf Acknowledgments}: We would like to thank Mr. K. Sankarasubramanian for his
constant help throughout the work, primarily in developing the algorithms and
implementing them with success. The services rendered by Messrs V. Chinnappan,
P. Anbazhagan, and V. Murthy are also greatfully acknowledged.
\vspace{0.5cm}

\begin{center} 
{\bf References} 
\end{center} 
\vspace{0.3cm}

\noindent
[1] A. Labeyrie, 1970, Astron. Astrophys., {\bf 6}, 85. \\
\noindent
[2] S. K. Saha, 1999a, Bull. Astron. Soc. Ind., {\bf 27}, 443.\\ 
\noindent
[3] S. K. Saha, 1999b, Ind. J. Phys., {\bf 73B}, 553.\\  
\noindent
[4] H. A., McAlister, 1988, Proc. ESO-NOAO conf. `High Resolution Imaging
Interferometry', ed. F. Merkle, Garching bei M\"unchen, FRG, 3.\\
\noindent
[5] G. Torres, R. P. Stefanik, and D. W. Latham, 1997, Astrophys. J, {\bf 485}, 
167.\\
\noindent
[6] W. I. Hartkopf, B. D. Mason, H. A. McAlister, N. H. Turner, D. J. Barry,
G. Franz, and C. M. Prieto, 1996, Astron. J, {\bf 111}, 936.\\
\noindent
[7] W. I. Hartkopf, H. A. McAlister, and B. D. Mason, 1997, CHARA Contrib. No. 
4, `Third Catalog of Interferometric Measurements of Binary Stars', W.I.\\
\noindent
[8] H. A. McAlister, W. I. Hartkopf, B. D. Mason, L. C. Roberts (Jr.),
and M. M. Shara, 1996, Astron. J, {\bf 112}, 1169.\\
\noindent
[9] B. D. Mason, D. R. Gies, W. I. Hartkopf, W. G. Bagnuolo (Jr.), T. ten 
Brummelaar, and H. A. McAlister, 1998, Astron. J, {\bf 115}, 821.\\
\noindent
[10] D. Bonneau, Y. Balega, A. Blazit, R. Foy, F. Vakili, and J. L. Vidal, 1986,
Astron. Astrophys. Suppl., {\bf 65}, 27.\\
\noindent
[11] A. Blazit, D. Bonneau, and R. Foy, 1987, Astron. Astrophys. Suppl., 
{\bf 71}, 57.\\
\noindent
[12] N. Miura, N. Baba, S. Isobe, M. Noguchi, and Y. Norimoto, 1992, J. Modern 
Opt., {\bf 39}, 1137.\\
\noindent
[13] I. I. Balega, Y. Y. Balega, I. N. Belkin, A. E. Maximov, V. G. Orlov,
E. A. Pluzhnik, Z. U. Shkhagosheva, and V. A. Vasyuk, 1994, Astron. Astrophys.
Suppl., {\bf 105}, 503.\\
\noindent
[14] E. P. Horch, Z. Ninkov, W. F. van Altena, R. D. Meyer, T. M. Girard, 
and J. G. Timothy, 1999, Astron. J, {\bf 117}, 548.\\
\noindent
[15] G. G. Douglass, R. B. Hindsley, and C. E. Worley, 1997, Astrophys. J 
Suppl., {\bf 111}, 289.\\
\noindent
[16] C. Leinert, A. Richichi, and M. Haas, 1997, Astron. Astrophys., {\bf 318},
472.\\
\noindent
[17] S. K. Saha, and P. Venkatakrishnan, 1997, Bull. Astron. Soc. Ind., 
{\bf 25}, 329.\\
\noindent
[18] M. E. Germain, G. G. Douglass, and C. E. Worley, 1999, Astron. J, 
{\bf 117}, 1905.\\
\noindent
[19] S. K. Saha, R. Sridharan, and K. Sankarasubramanian, 1999a, `Speckle image
reconstruction of binary stars', Presented at the Astron. Soc. Ind. meeting,
held at Bangalore, 1999.\\
\noindent
[20] Knox K.T. \& Thompson B.J., 1974, Ap.J. Lett., {\bf 193}, L45. \\ 
\noindent
[21] Lynds C.R., Worden S.P., Harvey J.W., 1976, Ap.J., {\bf 207}, 174. \\
\noindent
[22] Lohmann A.W., Weigelt G., Wirnitzer B., 1983, Appl. Opt., {\bf 22}, 4028.\\
\noindent
[23] Ayers G.R. \& Dainty J.C., 1988, Optics Letters, {\bf 13}, 547. \\
\noindent
[24] S. K. Saha, G. Sudheendra, A. Umesh Chandra, and V. Chinnappan, 1999b,
Experimental Astron., {\bf 9}, 39.\\
\noindent
[25] McAlister H. A. \& Hartkopf W. I., 1988, CHARA contribution no. 2. \\
\end{document}